\documentclass{camera}

\usepackage{graphicx}

\begin{document}

\title{An apparent GRBs evolution around us or a sampling of thin GRB beaming jets?}

\author{Daniele Fargion $^*$ , Daniele D'Armiento }
\organization{Physics Department Rome 1 La Sapienza and $^*$  INFN,
Italy;}

\maketitle

{\centerline{\bf Abstract} \small \vskip1mm \noindent

The gamma ray burst \emph{apparent average isotropic}  power versus their red-shift of all known GRB (Sept.2009) is reported. It calls for an
unrealistic Gamma Ray Burst Evolution around us or it just probe the need of a very thin gamma
precession-jet model.    These precessing and spinning $\gamma$ jet are originated by
Inverse Compton and-or Synchrotron Radiation at pulsars or
micro-quasars sources, by ultra-relativistic electrons. These Jets
are most powerful at Supernova birth, blazing, once on axis, to us
and flashing GRB detector. The trembling of the thin jet (spinning, precessing, bent by magnetic fields) explains
naturally the observed erratic multi-explosive structure of
different GRBs, as well as its rare re-brightening. The jets are precessing (by binary companion or
inner disk asymmetry) and decaying by power law $\frac{t_o}{t}$ on
time scales $t_o$ a few hours. GRB blazing  occurs inside the
observer cone of view only a seconds duration times; because relativistic synchrotron (or IC) laws the
jet angle is thinner in gamma but wider in X band. Its apparent brightening is so well correlated with its hardness (The Amati correlation). This explain the wider and longer X GRB afterglow duration and the (not so much) rare presence of
X-ray precursors well before the apparent main GRB explosion. The jet lepton maybe originated by an inner primary
hadron core (as well as pions and muons secondary Jets). The EGRET, AGILE and Fermi few  hardest and late GeV gamma might be PeV neutron beta decay in flight observed in-axis under a  relativistic  shrinkage.

\section{Introduction}
 The well probed Super-novae-GRBs connection since $1998$ naturally require  a thin beaming whose softer external cone
 (as for nearest GRB980425) is explaining the huge diversity between spherical SN output  and apparent coexisting GRB.
  Last and nearest GRB-XRF  080109 has been an exceptional lesson on GRB nature. Its lowest output maybe understood
  as the external tail of GRB jet. It also calls for a
  huge population of such SN-XRF in far Universe, undetected because below the present Swift,Fermi threshold.
   After a decade (since 25 April 08) we know by sure  that  Supernovae may often contain a Jet.
   Its persistent  activity  may shine on axis as a GRBs. Such a persistent,
  thin  beamed gamma jet may be powered by either a BH (Black Holes) or Pulsars.
  Late stages of these jets may loose the SN traces and appear
  as a short GRB or a long orphan GRB (depending on jet angular velocity and
  view angle). XRF are peripheries viewing of the same GRB inner jets.

  %%%%%%%%%%%%%%%%%%%%%%%%%%%%%%%%%%%%%%%%%%%%%%%%%%

\begin{figure}[h]
\begin{center}
\includegraphics[width=4in, height=2.0in]{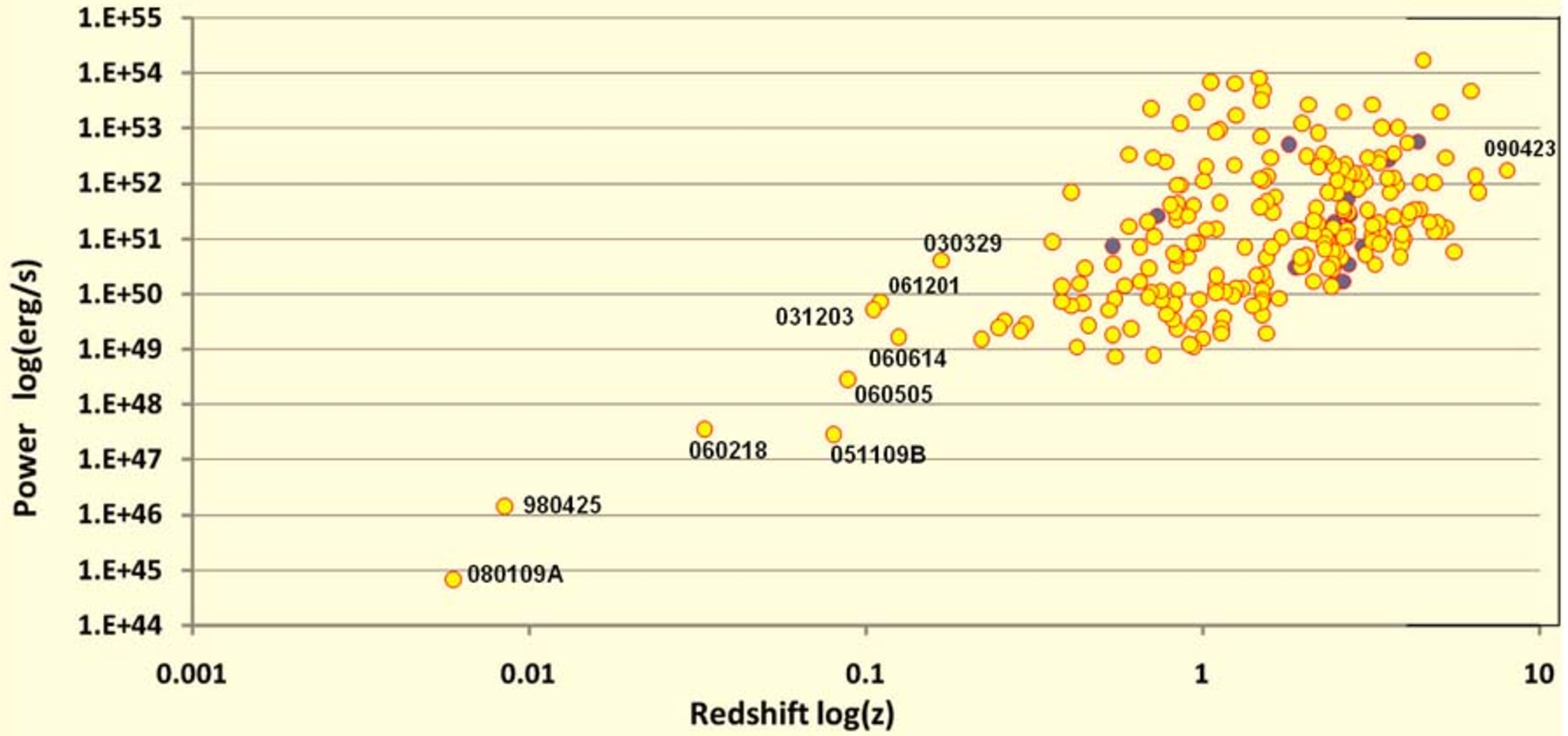}
 \caption{The GRB X-ray luminosity updated  to September 2009.
 The Luminosity vs red-shift law in a quadratic power, is mostly due (in lower regions) to the quadratic
 distance cut-off and (in higher regions) to the rarer beaming in axis occurring mostly by largest samples and cosmic volumes.
 The spread of nearly ten order of magnitude in luminosity (iso) calls for a thin ($0.001-0.0001$ rad and micro-nano-sr solid angle) beams.
  The new events 2009 are marked by a  gray dots. The key events are with their name label. }\label{Fig00}
\includegraphics[width=4in, height=2.0in]{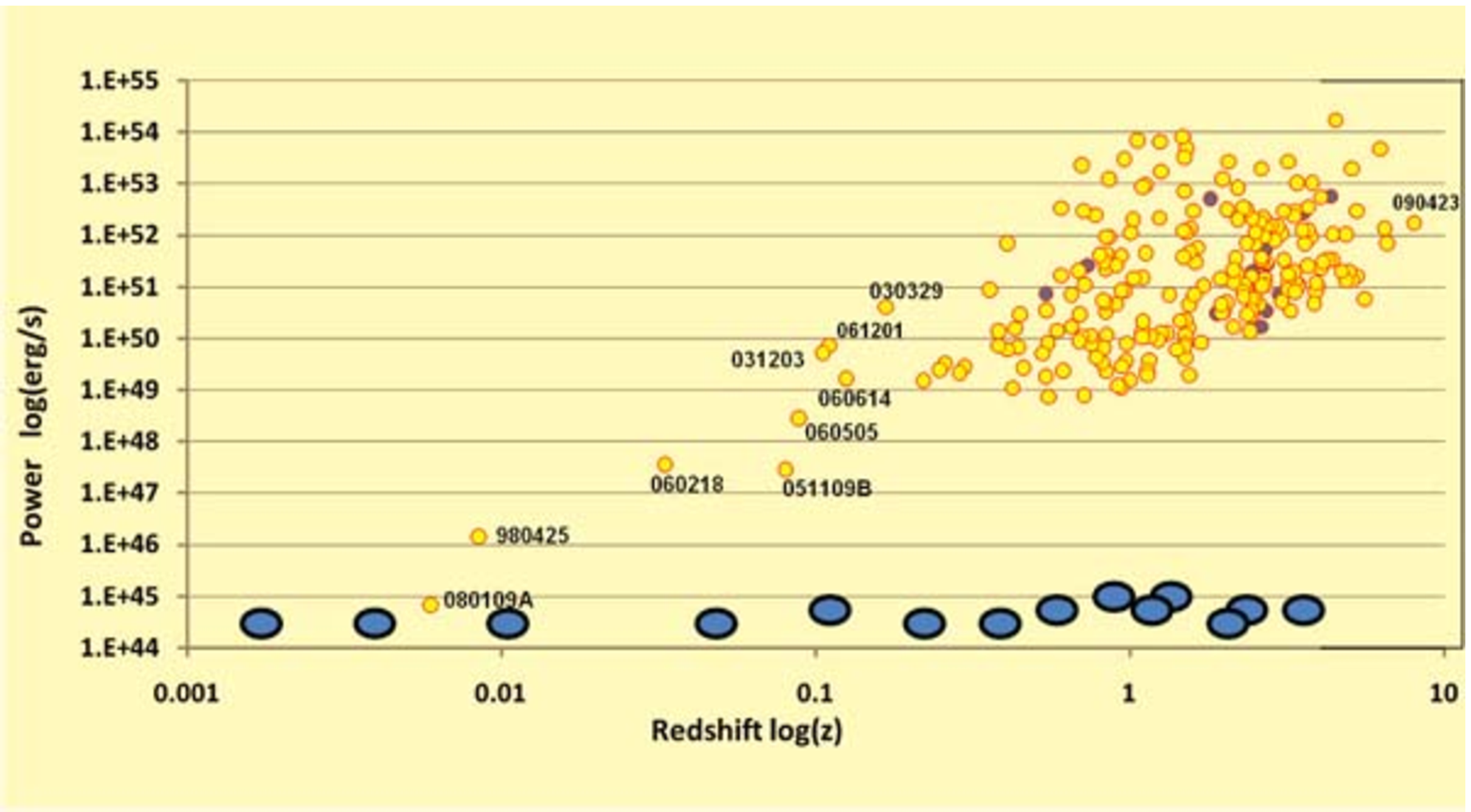}
 \caption{The GRB gamma and X-ray luminosity updated  to September 2009.
 The Luminosity vs red-shift observed for an imaginary standard candle Supernova isotropic explosions (circled rings below).  They differ from the \emph{apparent wide spread populations} of GRBs} \label{Fig01}
 \end{center}
\end{figure}

\begin{figure}[h]
\begin{center}
\includegraphics[width=4in, height=2.0in]{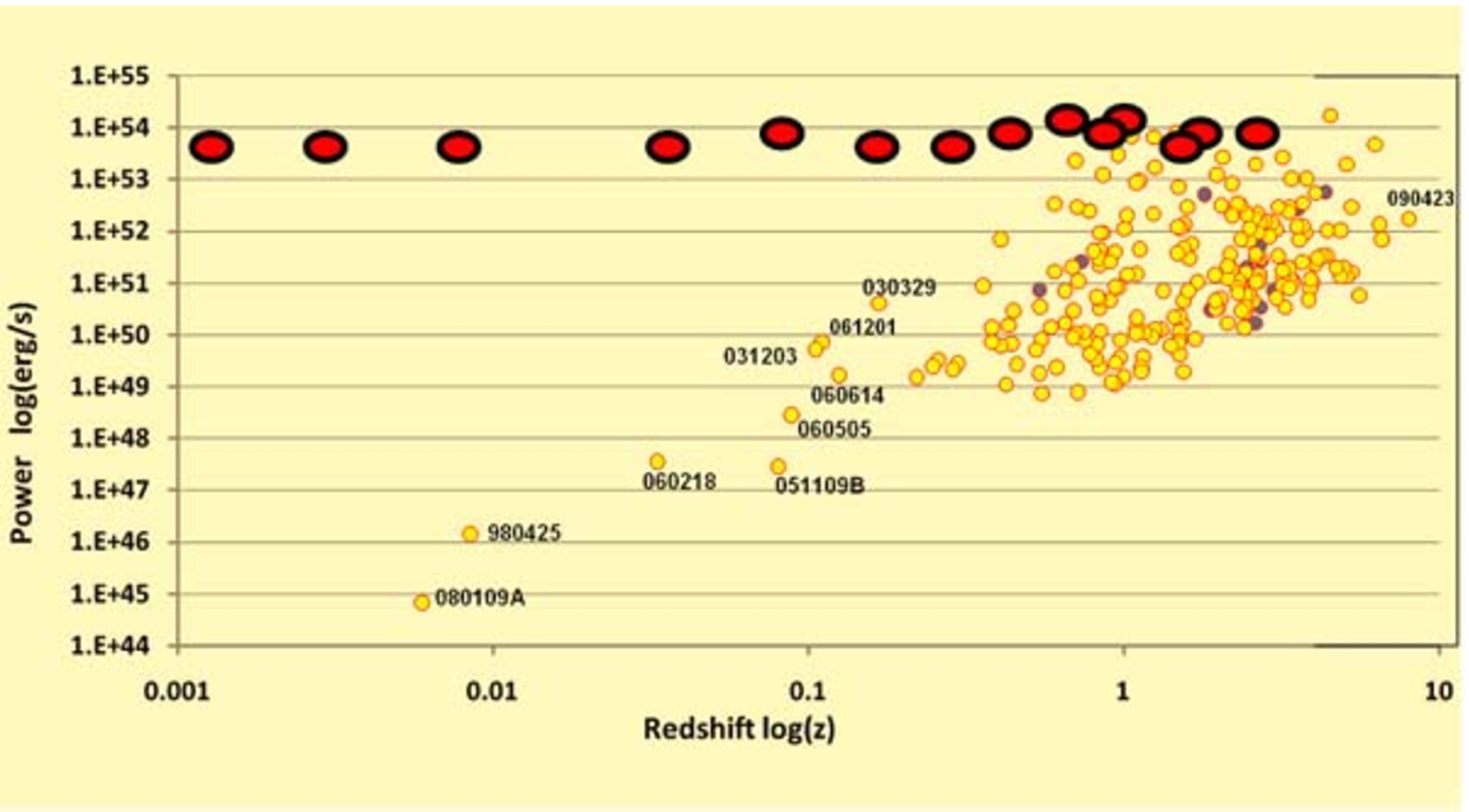}
 \caption{As above the hypothetical  GRB Fireball (isotropic) Luminosity vs red-shift strongly in disagreement  to the observed GRB one. }\label{Fig02}
\includegraphics[width=4in, height=2.0in]{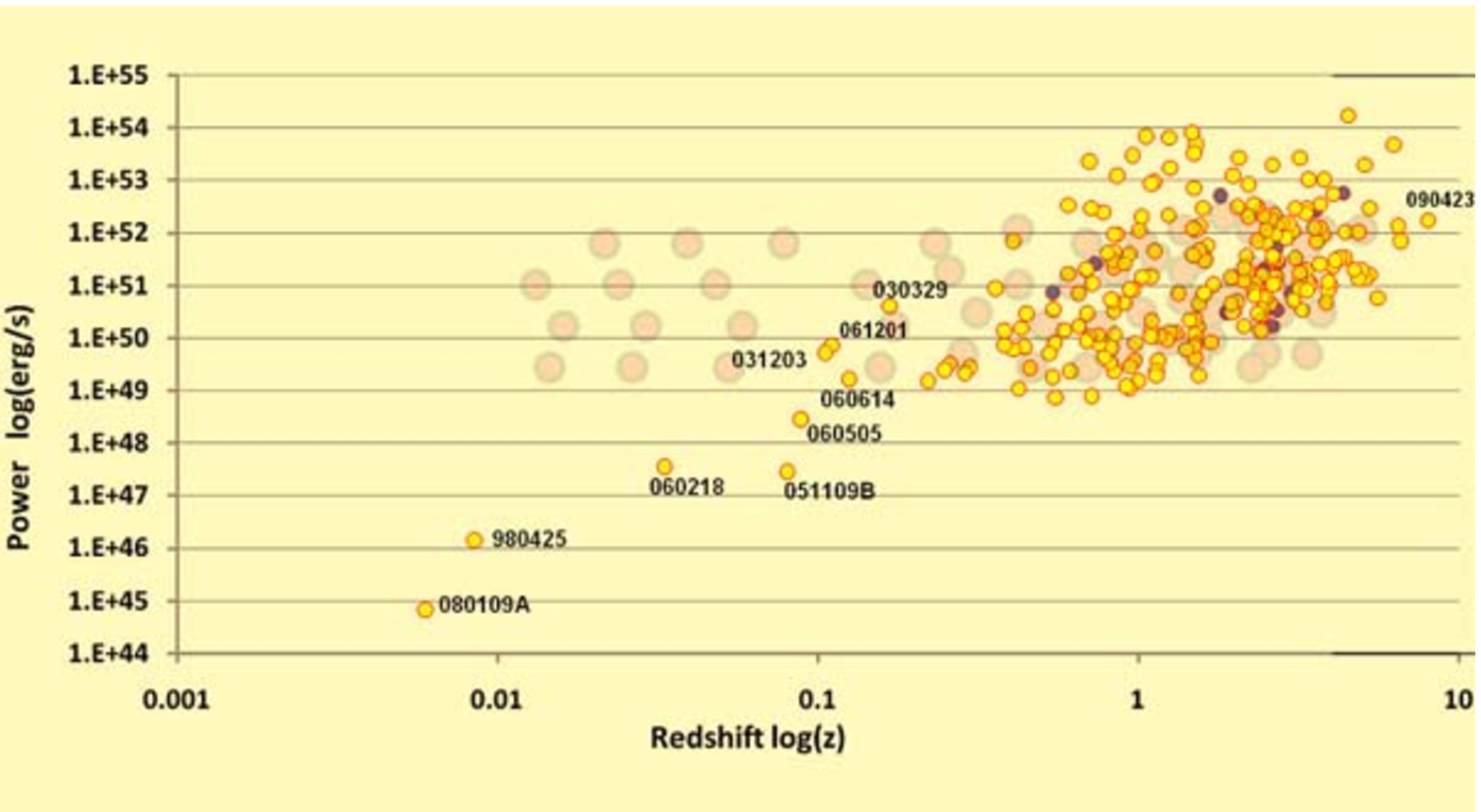}
 \caption{As above an hypothetical (popular) GRB Jet-Fountain Fireball (Jet $0.001$ sr solid cone), one shoot. The events are overlap on the record of data. The luminosity vs red-shift is partially in agreement (an illusion of success) only for redshift above unity, but it is still in strong  disagreement  with the   observed GRB luminosity at nearest ($z<<1$) distances.  }\label{Fig03}
\end{center}
\end{figure}

%%%%%%%%%%%%%%%%%%%%%%%%%%%%%%%%%%%%%%%%%%%%%%%%

\section{Actual unresolved GRB puzzles}

GRBs understanding (for Fireball one shoot models) is still an open challenge; many questions need
to be answered. The total energy output spans more than 8 order of
magnitude (\cite{Fa99}), with the most powerful and variable events
residing at the cosmic edges (\cite{Yo2004}), see Fig.\ref{fig1}.
This is apparently contrary to any reasonable Hubble law (\cite{Fa99}).
GRBs peak energy follow the so-called Amati correlation
\cite{Amati}. It has not a great cosmic meaning, but it simply correlate the geometry beaming
and the relativistic inner harder spectra. The most far ones are more abundant and more of them
are better beamed and more \emph{apparently} hard and bright.  This \emph{law} holds also within a single GRB event,
 between peak and peak activity while the jet, spinning and precessing, is bent in and out the target (our Earth).
When the jet is more aligned to the observer the apparent luminosity and hardness increase.
 There is another (somehow comparable) relation regarding the total energy
emitted versus redshift, which is far from being negligible, but it is less compelling
because it combine the whole event history. Moreover power is a relativistic invariant.The energy is not.

%%%%%%%%%%%%%%%%%%%%%%%%%%%%%%%%%%%%%%%%%%%%%%%%
\begin{figure}[h]
\begin{center}
\includegraphics[width=2.8in]{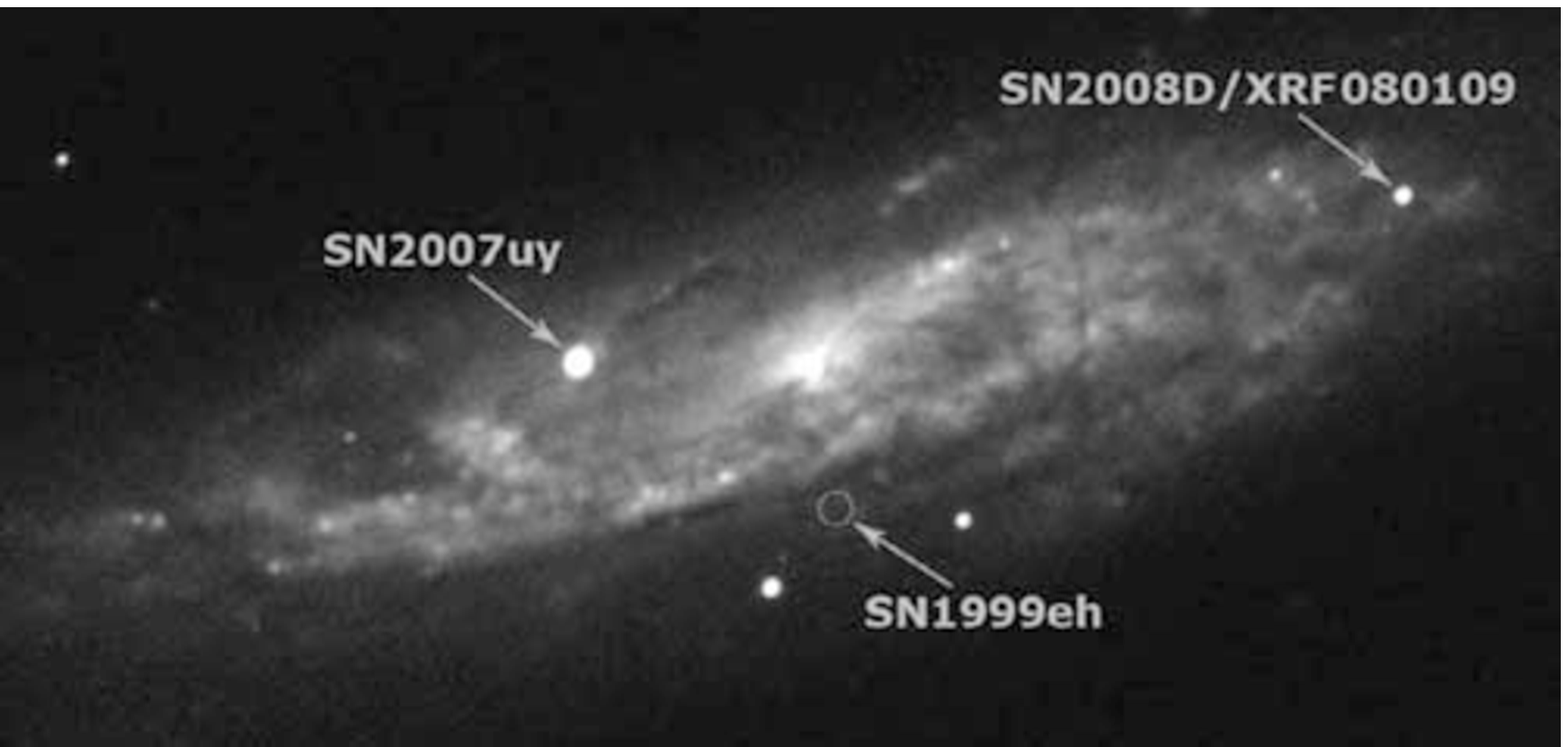}
\caption{The rare NGC 2770 twice SN within a week time: the XRF080109-SN2008D has deep meaning even for most sceptic theorist} \label{Fig04}
\includegraphics[width=3.5in]{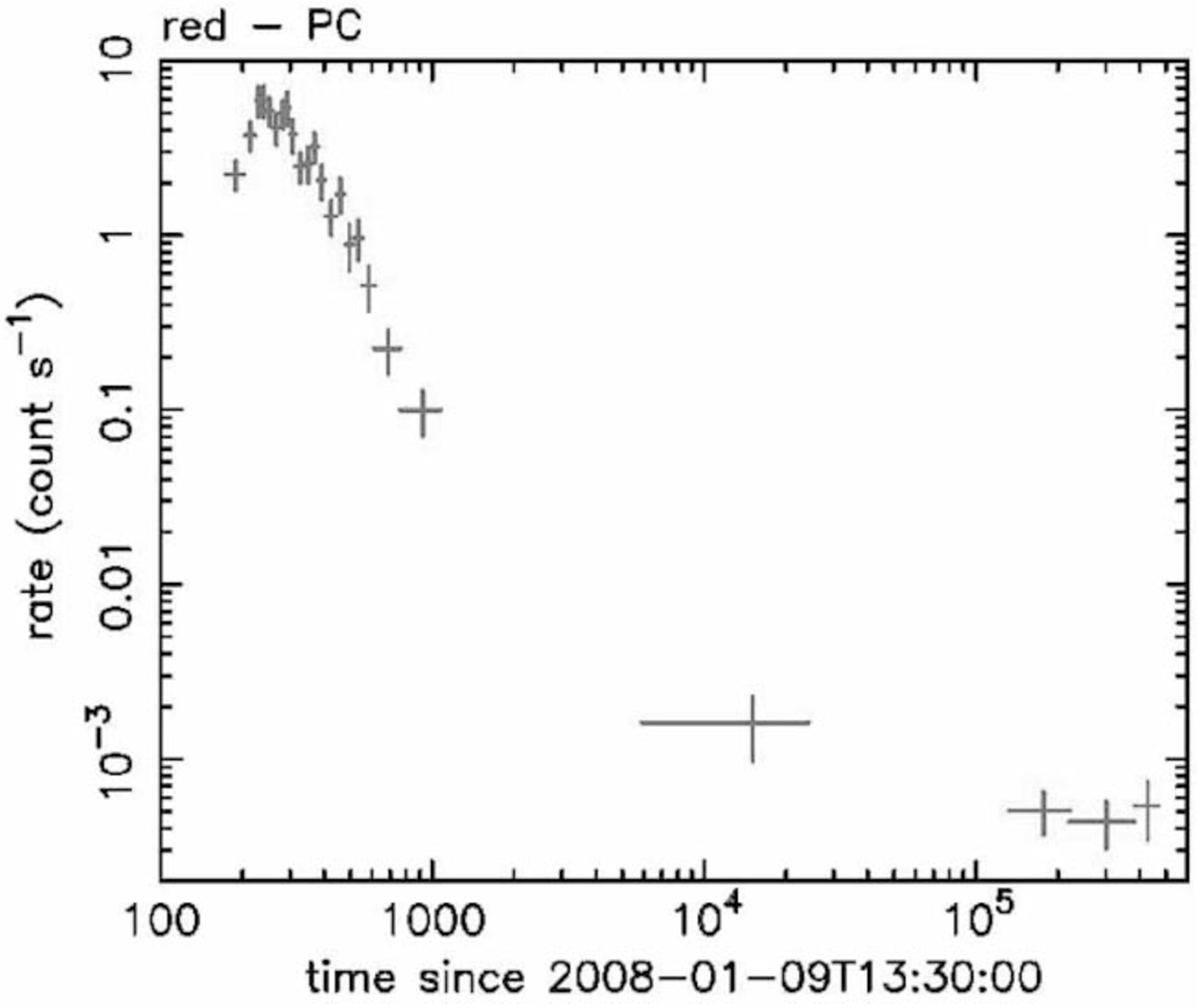}
 \caption{The above long XRF luminosity imply a new object or just a SN-GRB jet whose precessing is observed much off-axis, nearly at widest angle} \label{Fig05}
 \end{center}
\end{figure}

\begin{figure}[h]
\begin{center}
\includegraphics[width=2.4in]{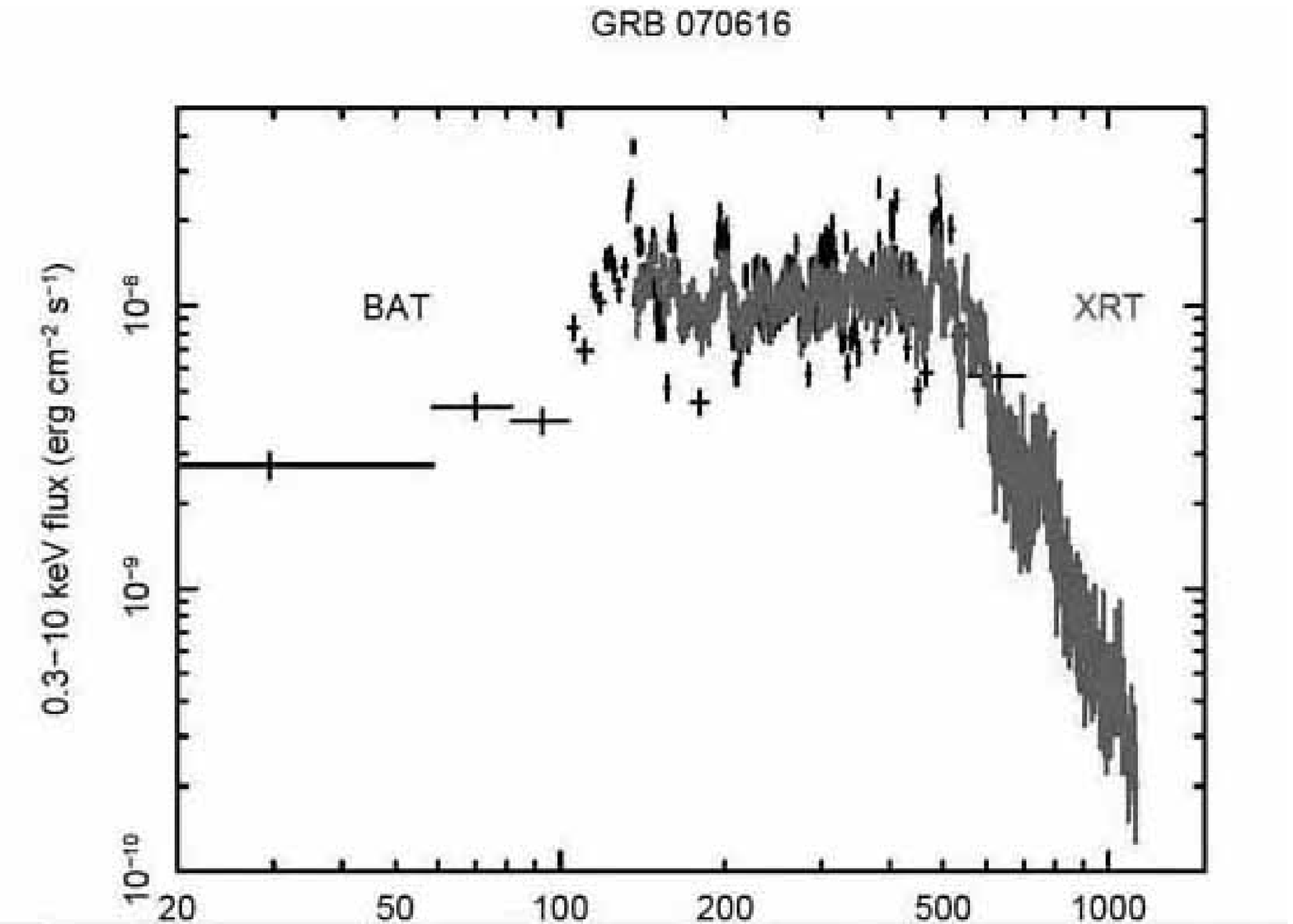}
\includegraphics[width=2.4in]{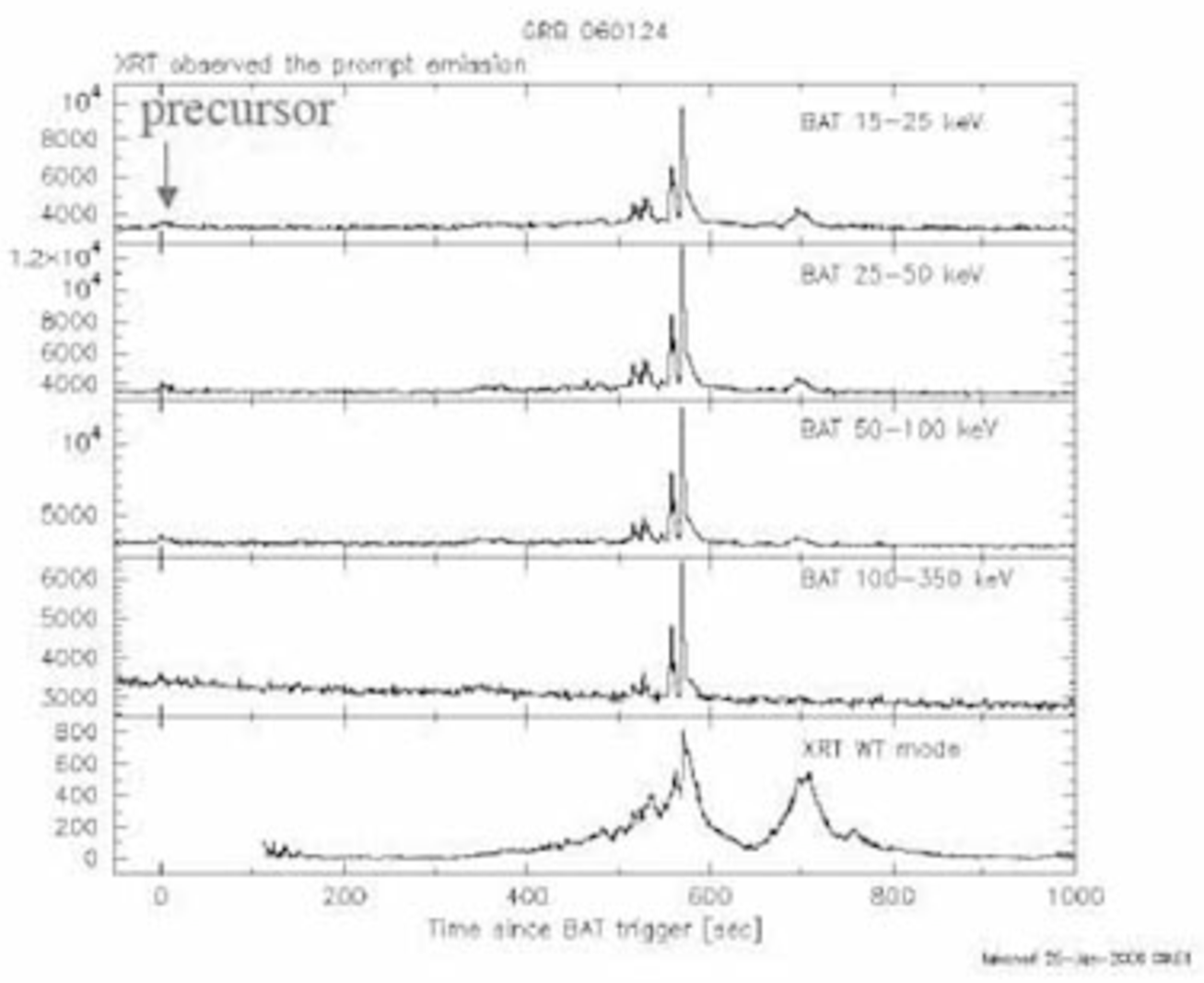}
 \caption{Left: The last GRB070616  long X-ray life; right: the puzzling ten-minute X-Ray precursor in GRB060124} \label{Fig06-Fig07}
\includegraphics[width=3in]{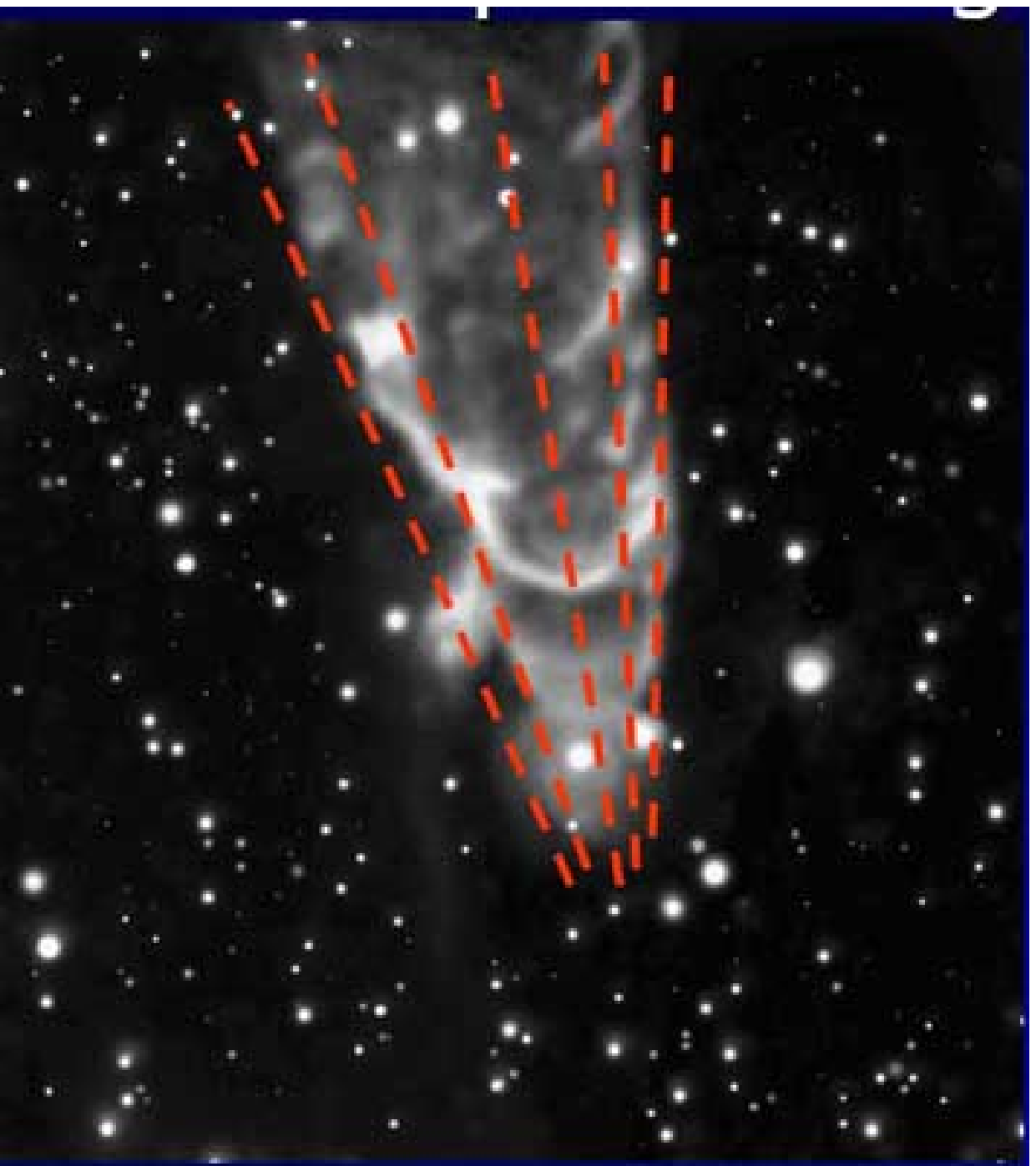}
\caption{From the left to the right: A possible \emph{3D} structure
view of the precessing jet obtained with a precessing and spinning,
gamma jet; In the panel we show an Herbig Haro-like object HH49, whose spiral
jets are describing, in our opinion, at a lower energy scale, such
precessing Jets as micro-quasars SS-433.} \label{Fig08}
\end{center}
\end{figure}

%%%%%%%%%%%%%%%%%%%%%%%%%%%%%%%%%%%%%%%%%%%%%%%%

Why the average GRB power is not a constant but it is a growing function (almost
quadratic) of the red-shift? Only a very few events are located in local universe, at lower
redshift (40-150 Mpc: just a part over a million of cosmic space),
such GRB980425 at $z=0.008$ and GRB060218 at $z=0.03$. Most
GRBs are located at largest distances, at $z\geq 1$ (\cite{Fa99}). Calling, apparently, to a rare
explosive event. However these nearby events have less power (also GRB030329), slow and
smooth evolution time respect to the farthest ones. Even if few statistically they (if isotropic)
occur more often than far ones.  Moreover they
show rough afterglows, bumps, re-brightening, such features being
very difficult to explain with one-shot explosive event (Fireball or Fireball-Jet). Such model
would predict monotonic decaying light curves, rather they often
show sudden re-brightening or bumpy afterglows at different time
scales and wavelengths (\cite{Stanek, DaF03}) - see e.g.
GRB050502B\cite{Falcone}.

Another relevant puzzling evidence (for Fireball and Magnetar) is the spectra structure similarity in few
GRBs and SGRs, hinting for the same origin rather than any beamed fountain
explosion and any isotropic magnetar (\cite{Fa99, Woo99}). Both should share a similar processes
(Thin precessing jet) but different distances and output.
Indeed, how can a \emph{popular jetted fireball} (with an opening angle of
$5^o$-$10^o$ and solid angle as wide as $0.1-0.01 sr.$) release an
energy-power $10^{50}$ erg$ s^{-1}$ , nearly 6 orders of magnitude
more energetic than $10^{44}$ erg$ s^{-1}$, the corresponding
isotropic SN? It is not explained why this enormous energy output imbalance
can occur same place at same time.
 \emph{Fireball Jet Model} need fine tune of multi-shells around a GRB in order to produce tuned shock explosions
and re-brightening with no opacity within minutes, hours, days
time-distances from the source(\cite{DaF03}), which is not realistic and
justified. A tiny, but still extremely powerful, precursor is followed, after ten minute of quite, by a huge
 more powerful explosion, such as in GRB060124 (at redshift
$z=2.3$): a 10 minutes precursor and subsequent bursts hundreds of
times brighter cannot be easily explained by Fireball, but it may be acceptable by
a persistent thin (decaying) jet, whose geometry beaming may flashes at different times and intensity.
Late (hundreds-thousands) GRB-SN event or BH jets   become by metamorphosis  SGR stages. Their huge flare are mostly
blazing flashes and not explosive revival. Indeed the SGR1806-20 of 2004 Dec. 27th, shows
no evidence of the loss of its period $P$ or its derivative
$\dot{P}$ after the huge \textit{so called Magnetar} eruption. In this
model its hypothetical magnetic energy reservoir (linearly
proportional to $P\cdot\dot{P}$) must be largely exhausted. In the later
 SGR1806 radio afterglows there is a mysterious two-bump radio
curve implying additional energy injection many days later SGR huge event. In
this connection the GRB021004 light curves (from X to radio) are
calling for an early and late energy injection. Also the SGR1806-20
polarization curve has been changing angle radically in short
($\sim$ days) timescale. We indebt them to a precessing jet blazing geometry. In similar way
the short GRB050724 was able to bump and re-bright a day after the main burst\cite{Campana}; the late energy contribute is comparable to the prompt one. In some sense it has been a repeater event (observed at different angle and time). About hardest
GeV components we remind that rarest EGRET GRB940217, at highest energetic events, could held more than $5000$s.
The GeV delayed tail, as discussed next paragraph, might be indebt to a hadron (as neutron decay in flight)
component of the jet. The main one is related to the electromagnetic-lepton Jet. This lepton jet maybe also the
secondary of an inner hadronic Jet core.

\begin{figure}[t]
\begin{center}
\includegraphics[width=3.2in]{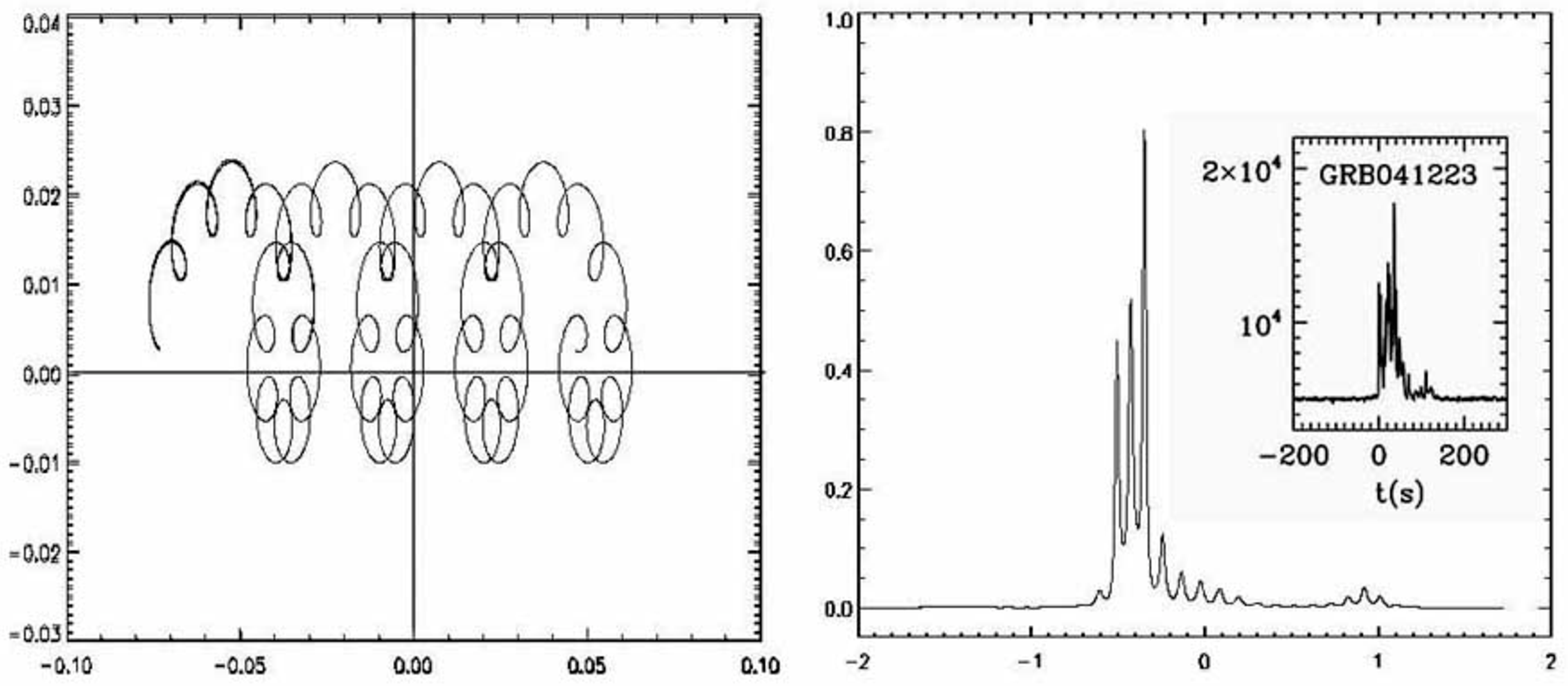}
\caption{The possible simple beam track of a precessing jet to
observer located at origin. On the left, observer stays in (0.00 ;
0.00); the progenitor electron pair jet (leading by IC\cite{FaSa98}
to a gamma jet) has here a Lorentz factor of a thousand and
consequent solid angle at $\sim\mu$ sr. Its consequent blazing light
curve corresponding to such a similar outcome observed in
GRB041223.}\label{Fig09}
\includegraphics[width=3.2in]{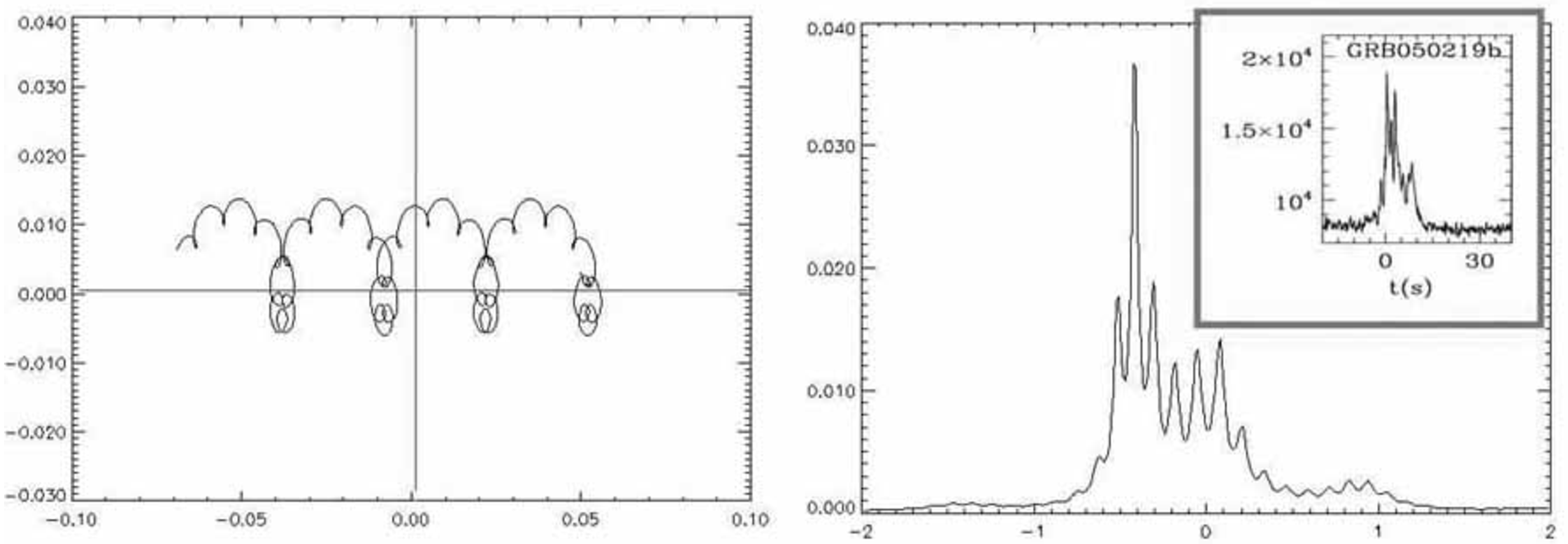}
\caption{Same as in Fig. \ref{Fig09}: a precessing jet and its
consequent light curve versus a similar outcome observed in
GRB050219b.} \label{Fig10}
\includegraphics[width=3.2in]{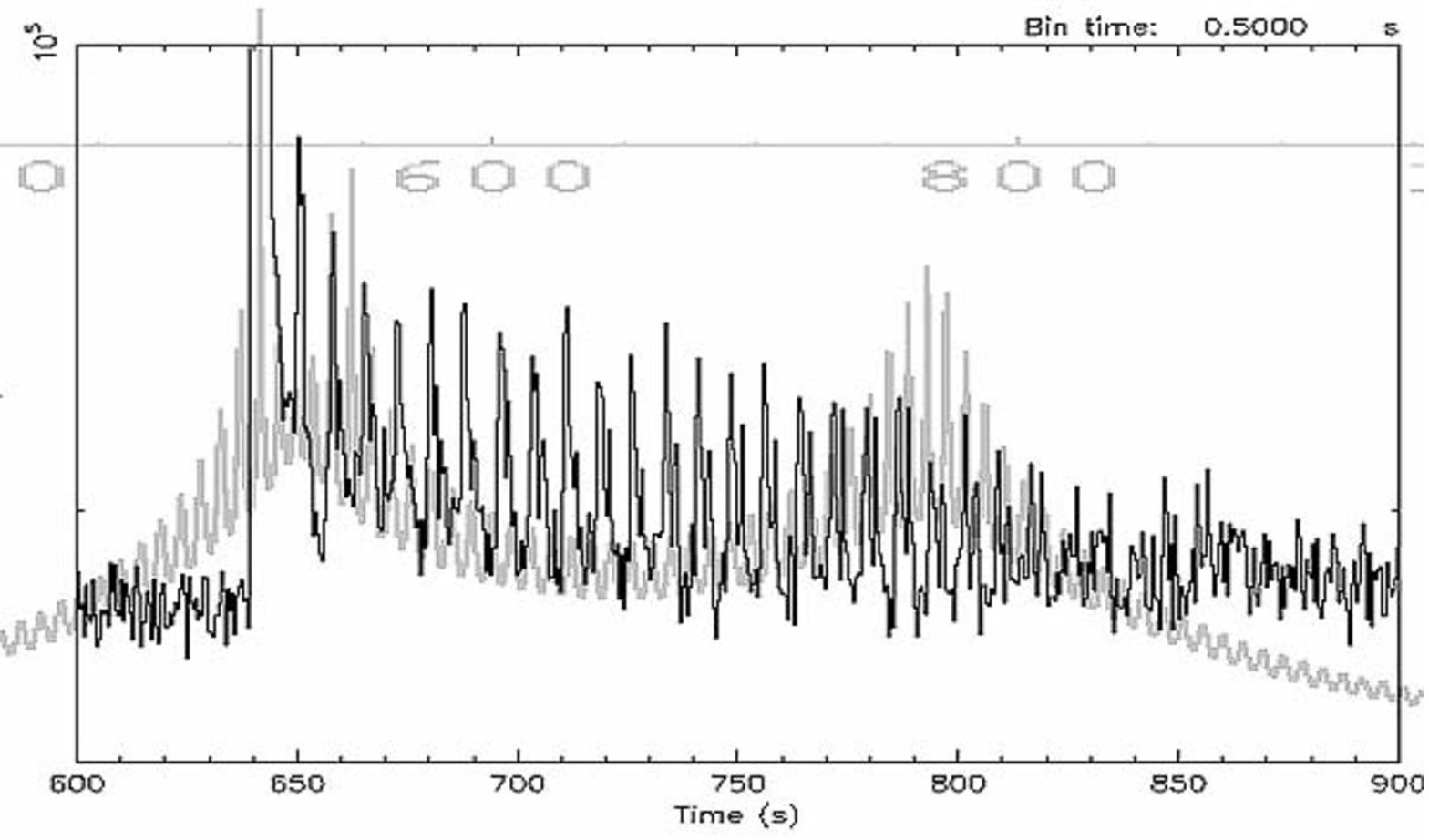}
\caption{Similar to Fig. \ref{Fig09}: a spiral, spinning precessing jet model by PeV muons, tens or hundred TeV electrons
radiating via synchrotron radiations and the consequent light curve (by persistent spinning and precessing jet) versus the similar outcome observed in huge flare
SGR1806-20.} \label{Fig11}
\end{center}
\end{figure}

Once these major questions are addressed and (in our opinion)
mostly solved by our precessing gamma jet model, a final  question
still remains, calling for a radical assumption on the thin
precessing gamma jet: how can an ultra-relativistic electron beam
(a common parental jet in any kind of Jet models) survive the SN background and dense
matter layers and escape in the outer space while remaining
collimated? Such questions are ignored in most Fireball fountain models
that try to fit the very different GRB afterglow light curves with
shock waves on tuned shells and polynomial ad-hoc curves around
the GRB event. The solution forces us more and more toward a
unified  precessing Gamma Jet model feeded by the PeV-TeV lepton
showering (about UHE showering beam see analogous ones\cite{Fa97,
Fa00-04}) into $\gamma$ discussed below. As we will show, the thin
gamma precessing jet is indeed made by a chain of primary
processes (PeV prompt hadrons and muon pair bundles decaying into electrons and then
radiating via synchrotron radiation), requiring an inner
ultra-relativistic jet inside the source.

%%%%%%%%%%%%%%%%%%%%%%%%%%%%%%%%%%%%%%%%%%%%%%%%%%%%%%%%%%%5

\section{PeV muons and neutrons bundles escaping from the GRB-SN explosion: the GeV delayed flare}
The spinning Jet maybe made by UHE (tens PeV) hadrons whose secondaries are muons bundles and later on electron pairs.
These pairs feed the gamma thin jet. However also neutrons tails (made by proton photo-pion conversions) at PeVs may escape the SN-GRB.
In this case their UHE neutron beta decay in flight nearly ( $t \simeq 30$ yrs) later maybe source of TeV electrons, themselves sources of tens GeV photons (by IR-TeV cut-off gamma tail showering). The neutron delay seem incomparable with observed delay (tens of seconds-minutes-hour). But these processes are \emph{apparently too late}. Indeed the decay of neutron occur on-axis and  it  is observed by relativistic shrinkage.
 A  relativistic reduction factor $$(1-\beta \cos\theta)\simeq \frac{\gamma^{-2}}{2}+ \frac{\vartheta^2}{2}$$ makes the neutron decay comparable with observed GeV delayed tail. The first factor for the neutron Lorentz boost (a million) is quite negligible, but for the second  a thin view angle $\theta\simeq 10^{-3}-10^{-4}$, ( just $10^{-7}-10^{-9}$ sr), fits the observed delay for the observed GRBs $$t' \simeq(1-\beta \cos\theta)t \simeq 100-1 s.$$. It should be noticed that both radiative (bremsstrahlung) beta decay and electron synchrotron radiation in galaxy field may offer GeVs photons.

\section{Blazing Spinning and
Precessing jets in GRBs}
 The huge GRBs luminosity (up to $10^{54}$
erg s$^{-1}$) may be due to a high collimated on-axis blazing jet,
powered by a Supernova output; the gamma jet is made by
relativistic synchrotron radiation and the inner the jet the
harder and the denser is its output. The harder the photon energy,
the thinner is the jet opening angle. The hardest and shortest
core Gamma event occur at maximal apparent luminosity once the jet
is beamed in inner axis. The jets whole lifetime, while decaying
in output, could survive as long as thousands of years, linking
huge GRB-SN jet apparent Luminosity to more modest SGR relic Jets
(at corresponding X-Ray pulsar output). Therefore long-life SGR
(linked to anomalous X-ray AXPs) may be repeating; if they are
around our galaxy they might be observed again as the few known
ones and the few rare extragalactic XRFs. The orientation of the
beam respect to the line of sight plays a key role in
differentiating the wide GRB morphology. The relativistic cone is
as small as the inverse of the electron progenitor Lorentz factor.
To observe the inner beamed GRB events, one needs the widest SN
sample and the largest cosmic volumes. Therefore the most far away
are usually the brightest. On the contrary, the nearest ones,
within tens Mpc distances, are mostly observable on the cone jet
\textit{periphery}, a bit off-axis. Their consequent large impact
crossing angle leads to longest \textit{anomalous} SN-GRB
duration, with lowest fluency and the softest spectra, as in
earliest GRB98425 and in particular recent GRB060218 signature. A
majority of GRB jet blazing much later (weeks, months after their
SN) may hide their progenitor explosive after-glow and therefore
they are called \textit{orphan} GRB. Conical shape of few nebulae
and the precessing jet of few known micro-quasar, describe in
space the model signature as well as famous Cygnus nebulae. Recent
outstanding episode of X-ray precursor, ten minutes before the
main GRB event, cannot be understood otherwise.

In our model to make GRB-SN in nearly energy equipartition the jet
must be very collimated $\frac{\Omega}{\Delta\Omega}\simeq
10^{8}$-$10^{10}$ (\cite{FaSa95b, Fa99, DaF05}) explaining why
apparent (but beamed) GRB luminosity $\dot{E}_{GR-jet}\simeq
10^{53}$-$10^{54}$ erg $s^{-1}$ coexist on the same place and
similar epochs with lower (isotropic) SN powers
$\dot{E}_{SN}\simeq 10^{44}-10^{45} erg s^{-1}$. In order to fit
the statistics between GRB-SN rates, the jet must have a decaying
activity ($\dot{L}\simeq (\frac{t}{t_o})^{-\alpha}$, $\alpha
\simeq 1$): it must survive not just for the observed GRB duration
but for a much longer timescale, possibly thousands of time longer
$t_o\simeq10^4\,s$. The late stages of the GRBs (within the same
decaying power law) would appear as a SGRs: indeed the same law
for GRB output at late time (thousand years) is still valid for
SGRs. SGRs are not Magnetar fire-ball explosion but blazing jets.
\section{The case of extreme SGR1806-20 flare and the GRB-SGR connection }
Indeed the puzzle (for one shot popular Magnetar-Fireball
model\cite{DuTh92}) arises for the surprising giant flare from SGR
1806-20 that occurred on 2004 December 27th: if it has been
radiated isotropically (as assumed by the Magnetar
model\cite{DuTh92}), most of - if not all - the magnetic energy
stored in the neutron star NS, should have been consumed at once.
This should have been reflected into sudden angular velocity loss
(and-or its derivative) which was \textit{never observed}. On the
contrary a thin collimated precessing jet $\dot{E}_{SGR-jet}\simeq
10^{36}$-$10^{38}$ erg $s^{-1}$, blazing on-axis, may be the
source of such an apparently (the inverse of the solid beam angle
$\frac{\Omega}{\Delta\Omega}\simeq10^{8}$-$10^{9}$) huge bursts
$\dot{E}_{SGR-Flare}\simeq10^{38}\cdot\frac{\Omega}{\Delta\Omega}\simeq10^{47}$
erg $s^{-1}$ with a moderate  steady jet output power (X-Pulsar,
SS433). This explains the absence of any variation in the
SGR1806-20 period and its time derivative, contrary to any obvious
correlation with the dipole energy loss law.

In our model, the temporal evolution of the angle between the
spinning (PSRs), precessing (binary, nutating) jet direction and
the rotational axis of the NS, can be expressed as
\[
\theta_1(t)=\sqrt{\theta_x^2+\theta_y^2}
\]
where
$$\theta_y(t)=
\theta_a\cdot\sin\omega_0t+\cos(\omega_bt+\phi_{b})+\theta_{psr}\cdot\cos(\omega_{psr}t+\phi_{psr})\cdot|(\sin(\omega_Nt+\phi_N))|+
$$
$$
+\theta_s\cdot\cos(\omega_st+\phi_{s})+\theta_N\cdot\cos(\omega_Nt+\phi_N))+\theta_y(0)
$$
and a similar law express the $\theta_x(t)$ evolution. The angular
velocities and phase labels are self-explained\cite{DaF05, DaF06}.
Lorentz factor $\gamma$ of the jet's relativistic particles, for
the most powerful SGR1806-20 event, and other parameters adopted
for the jet model represented in Fig. \ref{Fig11} are shown in the
following Table \ref{Tab1} (\cite{DaF05, DaF06}).
\begin{table}[h]
\begin{center}
\begin{tabular}{lll}
\hline \hline
  % after \\: \hline or \cline{col1-col2} \cline{col3-col4} ...
  $\gamma = 10^9$  & $\theta_a=0.2$ & $\omega_a =1.6 \cdot 10^{-8}$ rad/s\\
  $\theta_b=1$ &  $\theta_{psr}$=1.5 $\cdot 10^7$/$\gamma$ & $\theta_N$=$5 \cdot 10^7$/$\gamma$ \\
$\omega_b$=4.9 $\cdot 10^{-4}$ rad/s &  $\omega_{psr}$=0.83 rad/s
& $\omega_N $=1.38 $\cdot 10^{-2}$ rad/s \\
$\phi_{b}=2\pi - 0.44$ &$\phi_{psr}$=$\pi + \pi/4$ & $\phi_N$=3.5
$\pi/2 + \pi/3$ \\
$\phi_s \sim \phi_{psr}$ & $\theta_s$=1.5 $\cdot 10^6$/$\gamma$ & $\omega_s = 25$ rad/s \\
 \hline \hline
\end{tabular}
 \label{Tab1}
\end{center}
\end{table}

The simplest  way to produce the $\gamma$ emission  would be by IC
of GeVs electron pairs onto thermal infra-red photons. Also
electromagnetic showering of PeV electron pairs by synchrotron
emission in galactic fields, ($e^{\pm}$ from muon decay) may be
the progenitor of the $\gamma$ blazing jet. However, the main
difficulty for a jet of GeV electrons is that their propagation
through the SN radiation field is highly suppressed. UHE muons
($E_{\mu}\geq$ PeV) instead are characterized by a longer
interaction length either with the circum-stellar matter and the
radiation field, thus they have the advantage to avoid the opacity
of the star and escape the dense GRB-SN isotropic radiation field
\cite{DaF05, DaF06}. We propose that also the emission of SGRs is
due to a primary hadronic jet producing ultra relativistic
$e^{\pm}$ (1 - 10 PeV) from hundreds PeV pions,
$\pi\rightarrow\mu\rightarrow e$, (as well as EeV neutron decay in
flight): primary protons can be accelerated by the large magnetic
field of the NS up to EeV energy. The protons could in principle
emit directly soft gamma rays via synchrotron radiation with the
galactic magnetic field
($E_{\gamma}^p\simeq10(E_p/EeV)^2(B/2.5\cdot10^{-6}\,G)$ keV), but
the efficiency is poor because of the too small proton
cross-section, too long timescale of proton synchrotron
interactions. By interacting with the local galactic magnetic
field relativistic pair electrons lose energy via synchrotron
radiation:
$E_{\gamma}^{sync}\simeq4.2\cdot10^6(\frac{E_e}{5\cdot10^{15}\,eV})^2(\frac{B}{2.5\cdot10^{-6}\,G})\,eV$
with a characteristic timescale
$t^{sync}\simeq1.3\cdot10^{10}(\frac{E_{e}}{5\cdot10^{15}\,eV})^{-1}(\frac{B}{2.5\cdot10^{-6}\,G})^{-2}\,s$.
This mechanism would produce a few hundreds keV radiation as it is
observed in the intense $\gamma$-ray flare from SGR 1806-20.
The inner multi-precessing and spinning jet to the observer may lead to an apparent
resonant bumps as well a first huge flash, see Fig.\ref{Fig11}.

The Larmor radius is about two orders of magnitude smaller than
the synchrotron interaction length and this may imply that the
aperture of the showering jet is spread in a fan structure
\cite{Fa97, Fa00-04} by the magnetic field,
$\frac{R_L}{c}\simeq4.1\cdot10^{8}(\frac{E_{e}}{5\cdot10^{15}\,eV})(\frac{B}{2.5\cdot10^{-6}\,G})^{-1}\,s$.
Therefore the solid angle is here the inverse of the Lorentz
factor ($\sim$ nsr). In particular a thin
($\Delta\Omega\simeq10^{-9}$-$10^{-10}$ sr) precessing jet from a
pulsar may naturally explain the negligible variation of the spin
frequency $\nu=1/P$ after the giant flare ($\Delta\nu<10^{-5}$
Hz). Indeed it seems quite unlucky that a huge
($E_{Flare}\simeq5\cdot10^{46}$ erg) explosive event, as the
needed mini-fireball by a magnetar model\cite{DuTh92}, is not
leaving any trace in the rotational energy of the SGR 1806-20, $
E_{rot}=\frac{1}{2}I_{NS}\omega^2\simeq3.6\cdot10^{44}(\frac{P}{7.5\,s})^{-2}(\frac{I_{NS}}{10^{45}g\,cm^2})$
erg. The consequent fraction of energy lost after the flare is
severely bounded by observations:
$\frac{\Delta(E_{Rot})}{E_{Flare}}\leq10^{-6}$. More absurd in
Magnetar-explosive model is the evidence of a brief precursor
event (one-second SN output) taking place with no disturbance on
SGR1806-20 \textit{two minutes before} the hugest flare of 2004
Dec. 27th. The thin precessing Jet while being extremely
collimated (solid angle
$\frac{\Omega}{\Delta\Omega}\simeq10^{8}$-$10^{10}$
(\cite{FaSa95b, Fa99, DaF05, DaF06}) may blaze at different angles
within a wide energy range (inverse of
$\frac{\Omega}{\Delta\Omega}\simeq10^{8}$-$10^{10}$). The output
power may exceed $\simeq10^{8}$, explaining the extreme low
observed output in GRB980425 -an off-axis event-, the long late
off-axis gamma tail by  GRB060218\cite{Fargion-GNC}),  respect to
the on-axis and more distant GRB990123 (as well as GRB050904).

\section{ Conclusion}
 \emph{The GRBs are not the most powerful
explosions, but just the most  collimated ones.} They are within a mundane Supernova output power. Their birth rate
is comparable to the SN ones (a few a second in the observable
Universe), but their thin beaming ($10^{-8}$ sr) make them extremely
rare ($10^{-8} s^{-1}$ ) to observe, while pointing to us at their very birth (days-months after the SN birth).
This peculiar geometry tuning explain the wide spread of apparent power shown versus
 SN and GRBs as shown in Fig(\ref{Fig00},\ref{Fig01},\ref{Fig02},\ref{Fig03})
 The persistent precessing (in a slow decay of scale time of hours)
 and the moving beam span a wider angle with time and
 it encompass a larger solid angle increasing the rate by 3 order of
 magnitude, just to agree with the observed GRB rate ($10^{-5} s^{-1}$ ); after a few hours $\simeq 10^4 s.$
the beam may hit the Earth and appear as a GRB near coincident
with a SN. After months the optical SN is fade and the event is \emph{orphan}.
  The power law decay mode of the jet make it alive at a
smaller power days, months and year later, observable mainly at
nearer and middle distance as a Short GRB or (at its jet
periphery) as an XRF or in our galaxy as a SGRs.  The link with SN
is guaranteed in Long GRB. The presence of a huge population of active jets fit a
wide spectrum of GRB morphology \cite{Giovannelli}. The nearest
(tens-hundred Mpc) are observable mostly off-axis (because of
probability arguments) while the most distant ones are seen mostly on
axis (because threshold cut at lowest fluxes). Therefore the hardest are often at highest redshift.
But the IR cut-off makes this gamma bounded. Now in our Universe
thousands of GRBs are shining at SN peak power, but they are mostly pointing else
where. Only  nearly one a day might be blazing to us and captured at SWIFT, Agile, Fermi
threshold level. Thousand  of billions  are blazing (unobserved)
as SGRs in the Universe. Short GRBs as well SGRs are born in SNRs
location and might be revealed in nearby spaces. The GRB-SGRs
connection with XRay-Pulsars make a possible link to Anomalous X-Ray pulsar
jets recently observed in most X-gamma sources as the famous Crab.
The possible GRB-SGR link to X-gamma pulsar is a natural
possibility to be considered as a grand unification of the model.
Our prediction is that a lower threshold Fermi satellite will
induce a higher rate of GRBs both at nearer volumes (as GRB060218
and GRB 980425) and at largest red-shifts, where \emph{apparent} hardest, and brightest, event occurs. In a puzzling evolution frame.
Therefore the most probable source of GRB hardest neutrinos are the most distant GRBs at highest redshift, whose photons are often hidden by photon-IR-photon opacity. Therefore orphan GRB might be among the best UHE GRB neutrino sources. Making their discover more difficult and uncorrelated to GRB.

\end{document}